# Removal and recovery of ammonia from simulated wastewater using $Ti_3C_2T_x$ MXene in flow electrode capacitive deionization

Naqsh E. Mansoor[1], Luis A. Diaz[2], Christopher E. Shuck[3], Yury Gogotsi[3], Tedd E. Lister[2]✉ and David Estrada[1,2,4]✉

Flowing electrode capacitive deionization systems (FE-CDI) have recently garnered attention because of their ability to prevent cross contamination and operate in uninterrupted cycles ad infinitum. Typically, FE-CDI electrodes suffer from low conductivity, reducing deionization performance. Utilization of higher mass loadings to combat this leads to poor rheological properties. Herein, $Ti_3C_2T_x$ MXene was introduced as 1 mg mL$^{-1}$ slurry electrodes in an FE-CDI system for the removal and recovery of ammonia from simulated agricultural wastewater. The electrode performance was evaluated by operating the FE-CDI system with a feed solution of 500 mg L$^{-1}$ $NH_4Cl$ running in batch mode at a constant voltage of 1.20 and −1.20 V in charging and discharging modes, respectively. Despite the low loading, $Ti_3C_2T_x$ flowing electrodes showed markedly improved performance, achieving 60% ion removal efficiency in a saturation time of 115 min with an adsorption capacity of 460 mg g$^{-1}$. To understand the high adsorption performance of the electrodes, physiochemical and structural analysis was done via a variety of characterization techniques such as SEM, TEM, XRD, DLS, and Raman spectroscopy. Cyclic voltammetry and galvanostatic charge/discharge profiles were obtained to evaluate the electrochemical properties of the electrodes. The system proved to be an energy-saving technology by exhibiting a charge efficiency of 58–70% while operating at an energy consumption of 0.45 kWh kg$^{-1}$. A 92% regeneration efficiency showed that the electrodes were stable and suitable for long term and scalable usage. The results demonstrate that MXenes have the potential to improve the FE-CDI process for energy-efficient removal and recovery of ammonia.

## INTRODUCTION

Energy and water exist in a complex symbiotic relationship; energy production requires extensive use of water resources, while water remediation and purification processes exert a strain on energy sources. For example, nuclear and coal power plants require between 20 and 60 gallons of freshwater for every kilowatt-hour (kWh) of energy generated[1]. Correspondingly, water remediation and recovery efforts consume 2% of the total energy generated in the United States[2,3]. Environmental efficiency and economics encourage conservation of both resources. Water itself is an abundant resource, but only 2.5% is readily accessible as freshwater, and there are already countries that rely on remediation efforts to obtain their fresh water supply[4]. As the global population continues to rise, energy and water consumption will increase while existing natural sources will continue to be depleted. In its annual 2019 report, the International Energy Agency (IEA) predicts an 85% increase in energy-related water usage in the upcoming years[3]. In light of this projection, it is imperative that versatile, cost-effective, and energy-efficient water technologies are developed. Wastewater reclamation is one step in a multi-stage solution to solve the looming freshwater availability crisis, proving to be both environmentally relevant and fiscally smart, leading to both purified water and useful industrial byproducts.

Ammonia is one of the most common contaminants found in domestic and industrial wastewaters. Due to its importance in the manufacturing and agricultural industries, ammonia is one of the most industrially produced chemicals with over 14 million metric tons produced in the United States alone in 2019[5,6]. The production of ammonia requires copious amounts of both energy and water. While ammonia is undeniably a valuable building block in the modern food production, excessive exposure is a valid concern, due to its toxicity[6,7]. The release of ammonia-rich agricultural wastewater into aquatic ecosystems can cause eutrophication, leading to disruptions in delicate ecological balances[8]. Ammonia can also find its way into domestic water supplies as a natural degradation byproduct of nitrogenous organic matter[9]. Groundwater can have high ammonia levels in certain regions due to geological permeability[10]. Short and long-term exposure to ammonia and ammonium salts have been known to cause kidney damage, nervous system dysfunction, and lung edema[11,12]. Owing to these detrimental effects, the Clean Water Act (CWA) prohibits industrial and agricultural facilities from releasing nitrogen (as ammonia) rich wastewater to waterbodies[13]. Despite Federal and State level regulations, rural communities and vulnerable population groups in particular remain at risk for water quality standards violations due to a lack of infrastructure required by traditional wastewater treatment facilities[14]. Considerable effort has been devoted to developing efficient ammonia removal methodologies[15–20]. Despite such efforts, biological nitrification remains the most widespread method for treating ammonia wastewater[21]. Biological nitrification and other similar processes require heavy capital and infrastructure investments to be effective[22]. In addition, the process is not only cumbersome and slow, but it also wastes embedded ammonia, which is otherwise a valuable product. In pursuit of sustainability, product conservation directly translates to energy conservation. Hence, there remains a need for a water remediation technology that is energy, water, and resource efficient while requiring minimal capital investments.

[1]Micron School of Materials Science and Engineering, Boise State University, Boise, ID 83725, USA. [2]Idaho National Laboratory, Idaho Falls, ID 83401, USA. [3]A.J. Drexel Nanomaterials Institute and Department of Materials Science and Engineering, Drexel University, Philadelphia, PA 19104, USA. [4]Center for Advanced Energy Studies, Boise State University, Boise, ID 83725, USA. ✉email: tedd.lister@inl.gov; daveestrada@boisestate.edu

Prevalent water remediation technologies such as aerobic and anaerobic bio-treatment, multiple effect distillation, multistage flash distillation, and physical–chemical treatments require up to 6.60 kWh kg$^{-1}$ of direct energy input for the treatment of municipal wastewater[23]. Reverse osmosis (RO) is the most common physical treatment technique, and typically uses 3.0–4.7 kWh kg$^{-1}$ of energy[24,25]. Among purification technologies, capacitive deionization (CDI) has been a promising contender[26,27]. It uses a small voltage (~1.20 V) applied across two high surface area electrodes to induce charge separation. Similar to a supercapacitor, the charged ions are stored in the electric double layer of the electrodes. Even though there has been increasing research interest in CDI[28], the unavailability of adequate electrode materials is a technological bottleneck. Since voltage reversal (or removal), causes desorption of the immobilized ions, the technique can be used for purification as well as retrieval. Several studies report high water recovery (80–90%) and low energy utilization (0.6 kWh m$^{-3}$) for desalination using CDI[29–31].

As a result of an increased research activity around CDI[28], several different cell architectures have been developed including inverted-, hybrid-, ultrafiltration-, flow-by-, desalination battery, membrane-, flow-through-, cation intercalation desalination, and flow-electrode CDI (FE-CDI)[32–34]. Apart from FE-CDI, all of these CDI cell architectures utilize stationary electrodes and hence require an additional regeneration step for ion desorption, leading to non-continuous operation[32,35]. Additionally, the regeneration step can cause cross-contamination between the effluent streams, resulting in reduced water recovery[36]. This step negatively affects the fundamental motivators for CDI technology: cost, time, and energy efficiency. Due to simultaneous electrode regeneration ability, FE-CDI is a pioneering electrochemical technology that promises continuous and infinite remediation even for high concentration feed waters[37]. The adsorption capacity of the system is controlled by regulating the flow rate, channel design, and the nature and loading of the electrode material[38].

The flow electrode is an important component of the FE-CDI cell. The rheological properties and flowability of the electrode contribute to the adsorption capacity, stability, and cyclability of the system. Due to homogeneity, stable colloidal slurries perform better as flow electrodes. Carbon and its derivatives such as graphene sponges, graphene oxide, carbon nanotubes (CNTs), and various composites have been investigated as CDI electrodes[39]. Carbon serves as an excellent prototype material due to its high surface area, electrical conductivity, and electrochemical stability[39]. Previous experimental studies have determined the suitability of carbon and its derivate materials as CDI electrodes by exploring properties via microscopy imaging and porosity analysis[40,41]. While carbon materials have been shown to perform well in stationary electrode cell architectures[42], they suffer from insufficient electrical conductivity in slurry based flow-electrode systems[43]. In prior studies[44,45], and in our experimental experience, remediating the conductivity problem by increasing carbon content (>15 wt%) leads to clogged flow channels, halting operation of an FE-CDI system.

To overcome such limitations, we turn to MXenes, which are a class of two-dimensional (2D) transition metal carbides, nitrides and carbonitrides with the general formula $M_{n+1}X_nT_x$ where M is an early transition metal (Ti, V, Nb, etc.), X is carbon and/or nitrogen, $T_x$ represents the surface terminations (=O, –F, –Cl, and –OH), and $n = 1–4$[46]. MXenes are highly conductive, hydrophilic, form stable colloidal solutions, and their production can be easily scaled-up with no loss of properties[47,48]. MXenes have already been proposed as materials useful for environmental remediation, including heavy metal adsorption, pollutant adsorption, and desalination, amongst others[49–51]. $Ti_3C_2T_x$ was the first MXene discovered and is the most widely studied[52]. Furthermore, $Ti_3C_2T_x$ was shown to pose no ecological risk to aquatic ecosystems[53]. Recently this material has been applied to conventional CDI owing to high surface area and electrical properties[54–58].

Many studies have shown that $Ti_3C_2T_x$ is a promising pseudocapacitive negative electrode for supercapacitors[59–62]. Similarly, ideal materials for aqueous electrochemical energy storage should have high specific capacitance, charge efficiency, and electrochemical stability at water's electrolysis potential (~1.23 V)[63,64]. These features are also required in high performance CDI electrodes, suggesting that $Ti_3C_2T_x$ will be appropriate for CDI systems. Bao et al. demonstrated the use of aerogel-like $Ti_3C_2T_x$ MXene electrodes in a conventional desalination CDI cell to report an unprecedented salt adsorption capacity of 45 mg g$^{-1}$ [55]. In a recent publication, Ma et al. used binder free pristine $Ti_3C_2T_x$ films to achieve a salt adsorption capacity of 68 mg g$^{-1}$ [54]. However, conventional CDI cells consistently experience co-ion expulsion, which limits the amount of ions that can be adsorbed and reduces their charge efficiency[65]. High charge efficiency leads to low operating energy and the use of ion exchange membranes in FE-CDI cell reduces co-ion expulsion. However, the necessity of a regeneration step in conventional CDI invariably increases energy consumption[35,36]. For CDI to be considered an energy-efficient technology, it is necessary to account for the total operating energy, including regeneration.

This study aims to evaluate and understand the performance of $Ti_3C_2T_x$ flow electrodes in an FE-CDI system. While $Ti_3C_2T_x$ has been previously explored as a stationary electrode in conventional CDI architectures, this is the first reported investigation pertaining to its suitability in a flow electrode cell architecture[54,55,58]. We use a variety of characterization techniques to show that the MXene's unique set of properties, including high electrical conductivity, high negative zeta potential, layered structure, and the ability to form stable colloidal solutions render it highly desirable to be employed as flow electrodes[66]. Using de-ammonification of simulated wastewater as an example, we demonstrate the suitability of $Ti_3C_2T_x$ as a high-performance, low loading flow electrode in FE-CDI systems for water purification. A schematic of the FE-CDI system is shown in Fig. 1. With energy, water, and resource conservation, FE-CDI holds the promise to surpass limitations of preceding water remediation technologies including the widespread RO systems. This work opens avenues for further investigations of MXenes in FE-CDI technology for applications including but not limited to critical materials recovery, water softening, desalination, and selective nutrient removal.

## RESULTS

### Material characterization

X-ray powder diffraction (XRD) results shown in Fig. 2a confirm the successful etching of Aluminum (Al) from the $Ti_3AlC_2$ MAX phase. It can clearly be seen that the non-basal plane peak at $2\theta = 39°$ visible in the MAX phase XRD pattern disappears as a result of flake formation[52,67]. In addition, the shift in the (002) peak from 9.60° to 7.29° is indicative of the expansion of the *c*-lattice parameter due to successful etching[52,67]. In the MILD method, etching and delamination happen simultaneously, with delamination occurring through intercalation of water ($H_2O$), and lithium ions ($Li^+$). The ion removal mechanism in FE-CDI is dependent on the type of electrode material used[28]. Carbon electrodes operate through ion adsorption on the charged surface of the particles, while MXenes function by allowing ion insertion between the individual sheets[68]. Hence, the interlayer spacing has a pronounced effect on the charge storage and the ionic transport properties of MXenes[69]. The etched $Ti_3C_2T_x$ has a total interlayer spacing of 2.70 Å, due to the intercalated water and lithium ions coupled with successful etching. This results in the characteristic expanded structure of MXenes.

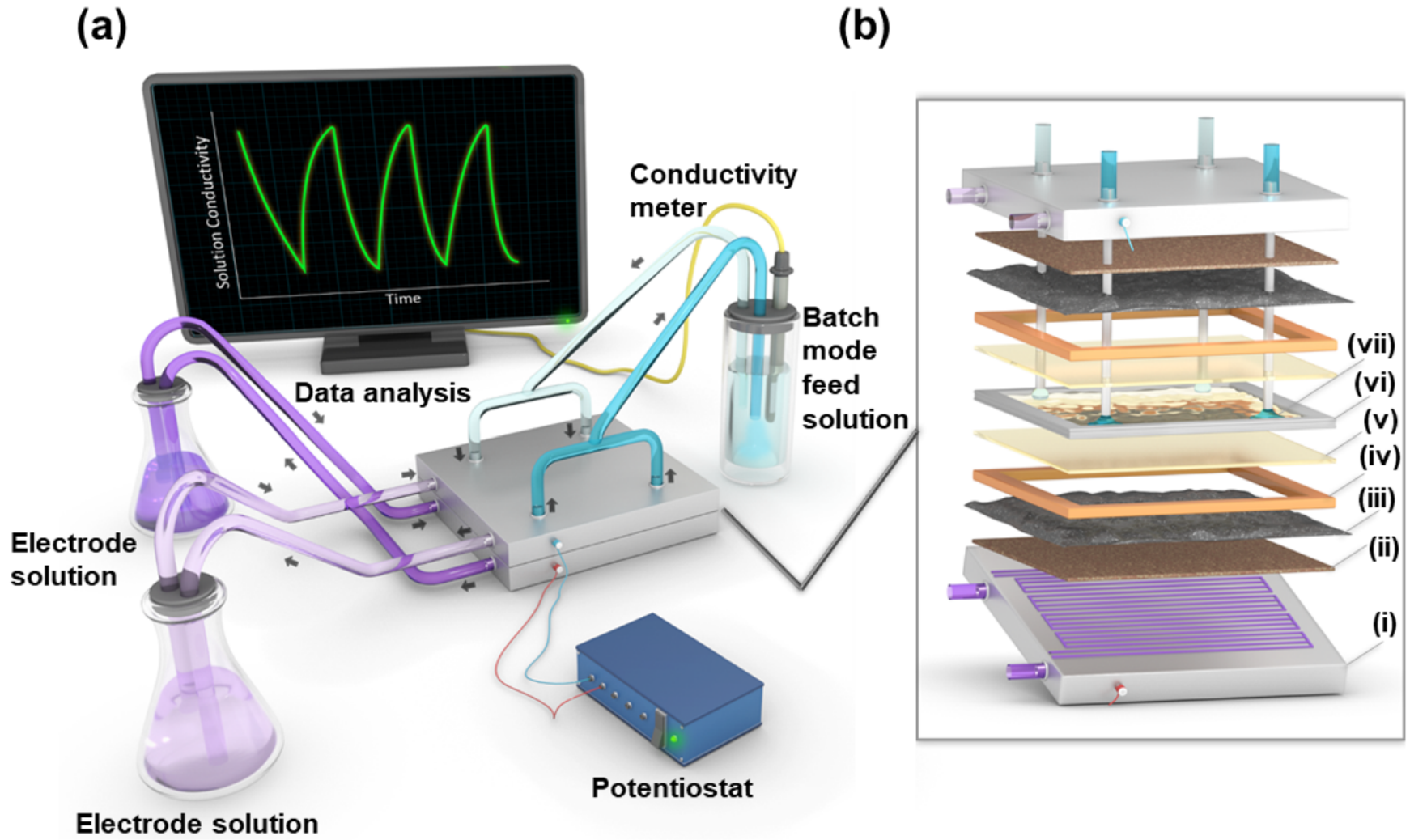

**Fig. 1 Schematic illustration of capacitive deionization operation.** Schematic illustration for **a** FE-CDI module for deionization testing, and **b** FE-CDI unit cell assembled with: (i) titanium current collectors; (ii) vitreous carbon; (iii) carbon cloth; (iv) rubber gaskets; (v) anion and cation exchange membranes; (vi) spacer; (vii) polyester filter felt.

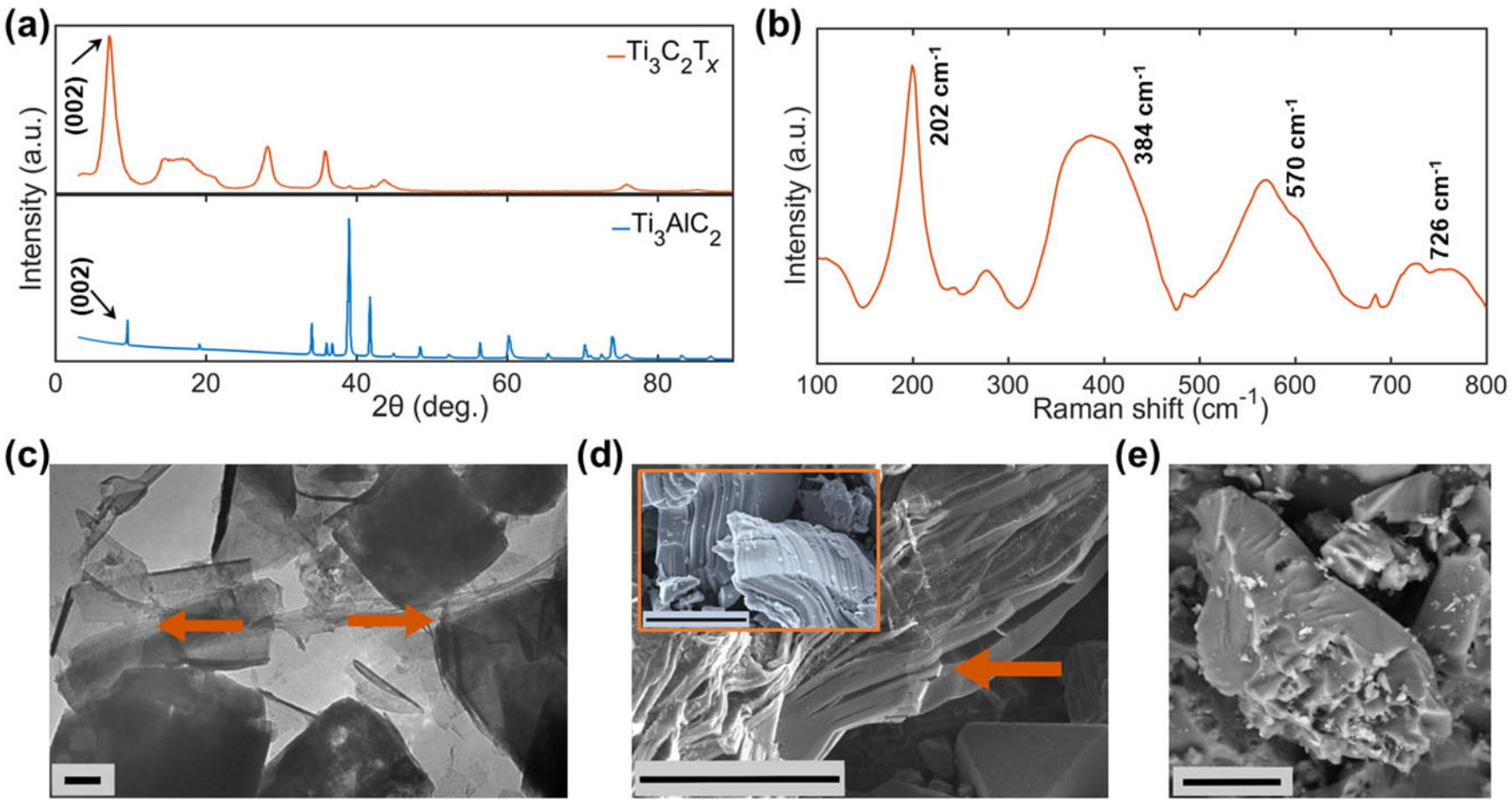

**Fig. 2 Materials characterization of FE-CDI electrodes.** **a** XRD patterns of $Ti_3AlC_2$ (bottom) and $Ti_3C_2T_x$ (top) before and after etching, respectively; **b** Raman spectra of $Ti_3C_2T_x$ flow electrode; **c** TEM image of as-etched $Ti_3C_2T_x$ with arrows highlighting separated layers; Scale bar = 0.2 µm. **d** SEM image of as-etched $Ti_3C_2T_x$ with SEM image of $Ti_3AlC_2$ (inset); Scale bar(s) = 5 µm. **e** SEM image of activated carbon (AC) powder particle. Scale bar = 5 µm.

The Raman spectrum of $Ti_3C_2T_x$ in Fig. 2b shows four distinct characteristic peaks at 202, 384, 570, and 726 $cm^{-1}$. The $A_{1g}$ peak at 202 $cm^{-1}$ and the $E_g$ peak at 384 $cm^{-1}$ correspond to vibrations due to surface groups on titanium[70]. The sharp and high intensity $A_{1g}$ peak at 202 $cm^{-1}$ indicates low defect flakes, as is expected from the chosen etching method[70]. The $E_g$ and $A_{1g}$ peaks observed at 570 and 726 $cm^{-1}$, respectively, can be attributed to carbon vibrations[70]. The relatively low intensity $E_g$ peak at 279 $cm^{-1}$ corresponds to the presence of –OH termination groups on the MXene surface and is a characteristic feature of the MILD synthesis method[52,70]. Line broadening and merging in the spectra is indicative of exfoliation and delamination and is hence consistent with the XRD data. Raman spectroscopic investigation of activated carbon (AC) to reveal structural information has been limited in its scope as the initial spectra produces broad and obscure peaks[71]. The initial relative D and G band intensities are insufficient to characterize the structural order of the material[71,72]. Since AC is desirable for its high surface area[73], other characterization techniques such as scanning electron microscopy (SEM), porosity, and particle size analysis are more informative.

The transmission electron microscopy (TEM) image in Fig. 2c shows stacked multilayer MXene sheets that are thin and electron transparent. The morphology and surface structure of the FE-CDI electrode particle materials have a significant effect on ion adsorption capacity. The highly accessible MXene surfaces, characterized by the expanded and open interlayer structure, allow for rapid ion adsorption within the MXene sheets[74]. The characteristic fanned out basal planes of etched MXenes can be seen in Fig. 2d. Moreover, the unfurled morphology is evidence for successful etching of $Ti_3AlC_2$ (Fig. 2d inset) and is hence in agreement with the aforementioned XRD and Raman results. It can be witnessed visually that the spread out, open structure of MXenes has significantly more intercalating space than the porous structure of AC (Fig. 2e). The AC also exhibits an irregular block morphology, with particle size ranging in a few microns.

## Flow electrode slurry characterization

Figure 3a shows particle size profiles obtained via DLS measurements of the 10 wt% AC slurry and the $Ti_3C_2T_x$ electrodes.

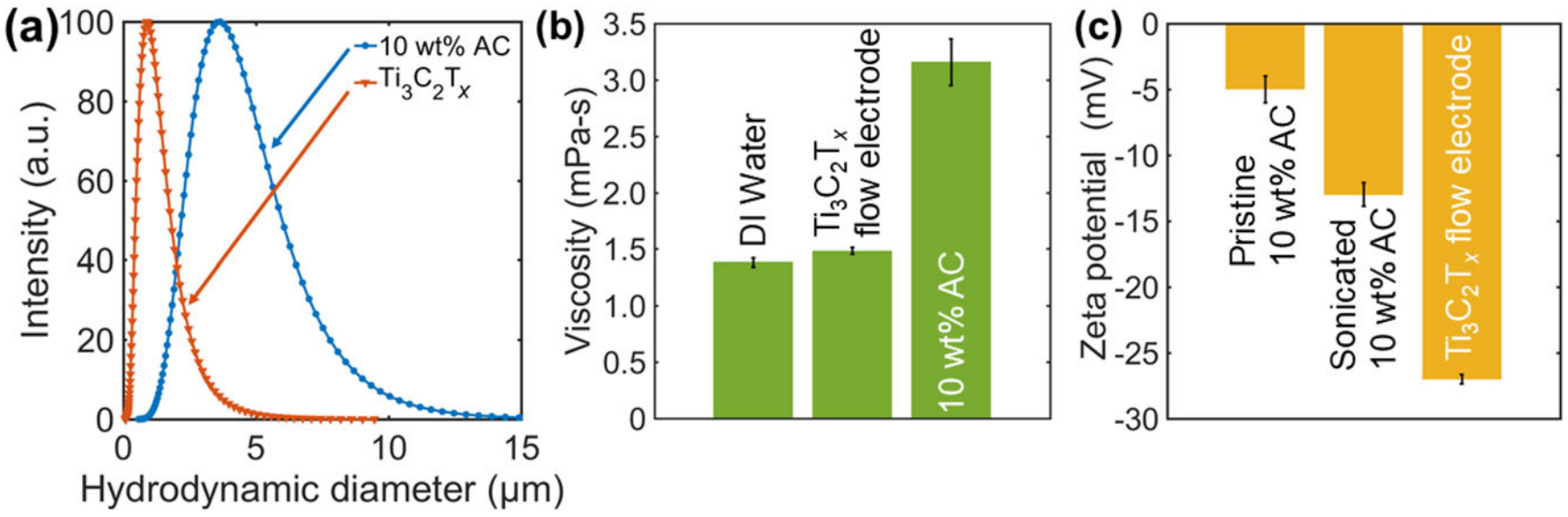

**Fig. 3 Rheological properties of FE-CDI electrodes. a** Particle size analysis of 10 wt% AC slurry electrode and $Ti_3C_2T_x$ flow electrode; **b** Viscosity measurements of DI water, $Ti_3C_2T_x$ flow electrode, and 10 wt% AC slurry electrode respectively; **c** Zeta (ζ) potential measurements of 10 wt% AC slurry before sonication, 10 wt% AC electrode after sonication, and $Ti_3C_2T_x$ flow electrode, respectively. Error bars represent standard deviation.

It is evident that the average particle size is lower for MXene electrodes (1.20 μm) than AC (4.50 μm). It is well established in the Rayleigh–Gans–Debye (RGD) theory that aggregates and large colloidal particles result in peak broadening in light scattering spectra[75–77]. The AC profile shown in Fig. 3a exhibits a wider distribution and longer tail end, hence indicating particle flocculation. Higher molecular weight flocculated particles are more prone to sedimentation which can lead to uneven electrode flow[78]. This was mediated to some degree by sonicating the AC slurry to reduce particle size. However, the AC slurry still required continuous stirring during the experiment to prevent sedimentation.

Furthermore, the viscosity of a suspension has a strong correlation with system performance. Higher viscosity contributes to poor flowability and dispersion of the slurry. Figure 3b shows that the $Ti_3C_2T_x$ solution has a viscosity very close to deionized (DI) water (1.50 and 1.40 mPa s, respectively). Experimentally, this resulted in excellent flowability and zero hindrance during the cell operation. As previously discussed, the MILD method results in the presence of hydrophilic functional groups on the surface of MXene layers which result in electrostatic repulsion that leads to a stable colloidal solution not prone to flocculation[79–81]. The presence of the hydrophilic surface terminations was confirmed by observing the surface group region (230–470 $cm^{-1}$) in the Raman spectra (Fig. 2b). Comparatively, the AC slurry had a higher viscosity (3.20 mPa s) and larger particle size (Fig. 3a) resulting in poor flowability and frequent clogging of the narrow cell channels in our conductor design. Zeta (ζ) potential is a quantitative measure of the electrical potential in the interfacial double layer and is an important guide to determine suspension stability[82]. Low (more negative) ζ-potential values indicate a lower probability of the suspension forming aggregates[83]. AC forms lyophobic colloids, which is reflected in its high (less negative) ζ-potential value of −5 mV. ζ-potential is affected by many factors including temperature, viscosity, and particle size[84]. To illustrate this, pristine and sonicated samples of AC slurry were compared against the $Ti_3C_2T_x$ flow electrodes for ζ-potential measurements as shown in Fig. 3c. As-received pristine 80 mesh AC powder was compared against the solution that was probe sonicated for 1 h to yield the particle size distribution given in Fig. 3a. It can be seen that particle size affects the ζ-potential between the two AC samples (−5 vs. −13 mV). In addition, the reduced particle size coupled with continuous stirring had a positive impact on fluid flow, as the cell channels were less likely to get clogged. As shown in Fig. 3c, the $Ti_3C_2T_x$ solution had a considerably lower (more negative) ζ-potential of −27 mV, which resulted in electrode stability and better flowability throughout the FE-CDI operation. This eliminated the need for any additional measures such as solution stirring to ensure an uninterrupted system operation.

### Deionization performance test

Figure 4a shows the change in conductivity ratio of the effluent solution as a function of time when the system was operated in batch mode. The observed cyclic conductivity change is representative of the ion capture and release steps during the regenerating operation. It is imperative to note that effluent conductivity change ($C_f/C_o$) and current response of the FE-CDI cell system are primarily a result of the properties of the electrode slurries. In the absence of any electrode system, using DI water in place of the flowing slurries, the baseline cell measurement shows insignificant current response and effluent conductivity change (Supplementary Figs. 1 and 2). In this study, the 10 wt% AC slurry was used as control electrodes to evaluate the performance of $Ti_3C_2T_x$. The conductivities of both deionization systems show a significant decrease with the application of 1.20 V external voltage. The open structure and intercalation capture mechanism in $Ti_3C_2T_x$ resulted in a shorter saturation time of 115 min compared to AC (233 min), and saturation times previously reported in other studies[41,85,86]. A similar phenomenon is observed when MXenes and other 2D materials are used for energy storage applications. The layered morphology and multiple storage sites lead to fast storage kinetics and high cycling rates[87–89]. Due to the shorter charge–discharge times, $Ti_3C_2T_x$ delivered 10 stable long-term cycles during the ~30 h run time. This manifested in twice the number of cycles as AC (5 cycles). As saturated electrosorption is achieved, the charging (ion capture) profile for $Ti_3C_2T_x$ plateaus at an average conductivity ratio of 0.410 ± 0.002, which is lower than the obtained value for AC (0.500 ± 0.001). It is interesting to note that the conductivity ratio of the first run in cycle with $Ti_3C_2T_x$ is markedly lower (0.250) than the following cycles. In the first cycle, the system has not yet achieved dynamic equilibrium and the increased deionization can result from permanent chemical interactions on the defect sites on the MXene flakes[90]. These ions are not desorbed upon voltage reversal (or removal). This is further evident by the fact that the subsequent regeneration cycle does not reach the initial conductivity ratio of 1. Hence, the first cycle is not representative of the electrode performance. This is reflected in Fig. 4b which shows electrode regeneration efficiency ($\eta_r$). After achieving dynamic equilibrium in the second cycle ($Ti_3C_2T_x$ after stabilization), $\eta_r$ is upwards of 92%. If we take the first cycle into account, $\eta_r$ drops to 69% which is evident of the suggestion that after the initial cycle, some adsorption sites are permanently occupied by chemical interactions. It is imperative to note that $\eta_r$ for AC is slightly higher at 96%. This can be attributed to the oxidative and aqueous degradation of $Ti_3C_2T_x$ over time[91]. However, researchers have been working to increase the oxidative stability of MXenes[92,93] so that this problem can be mitigated by employing environmentally stable MXenes. In addition, the use of MXenes in non-aqueous solvents as flow electrodes can also be explored in the future[94].

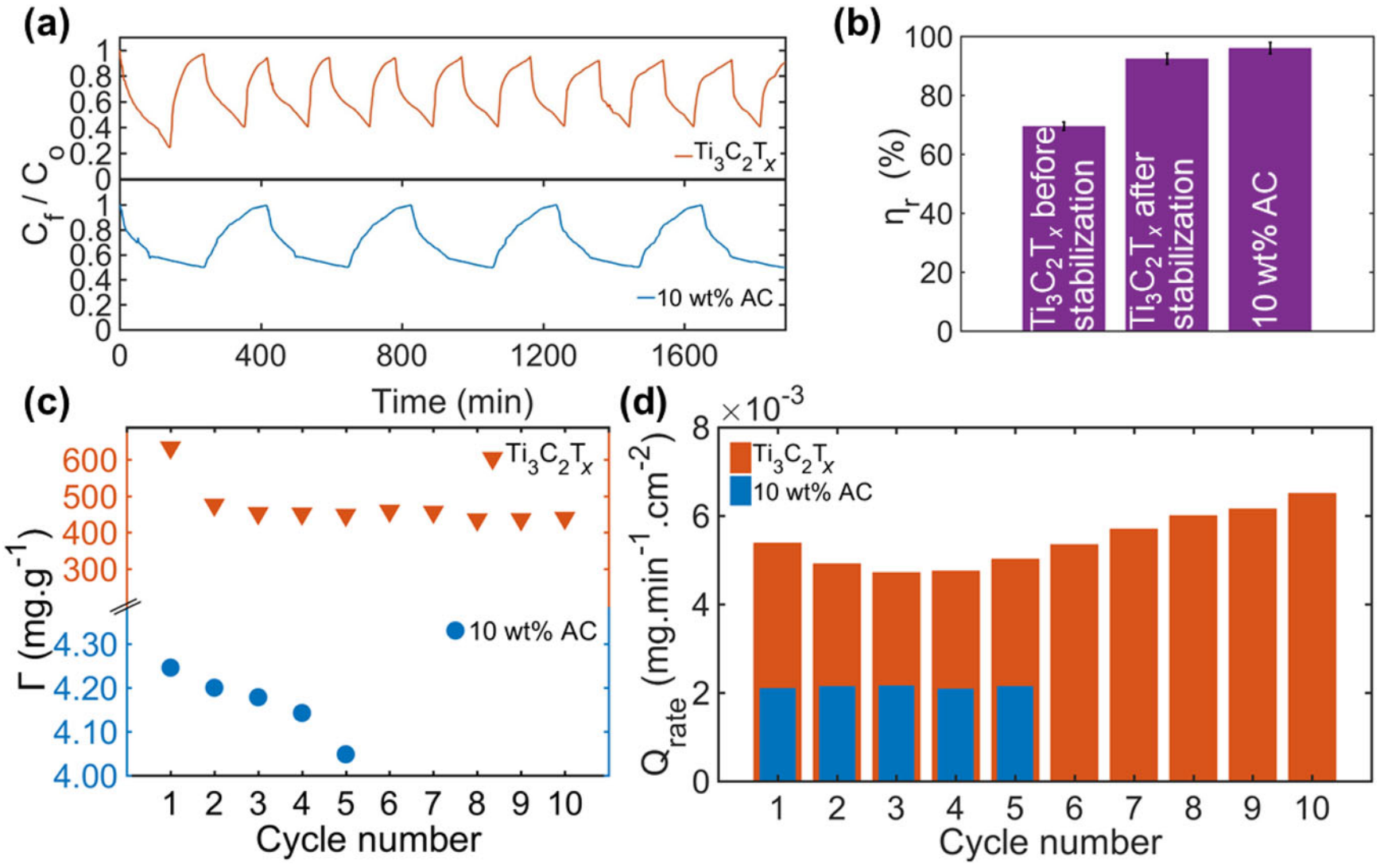

**Fig. 4 Electrode(s) performance during deionization testing. a** Effluent conductivity ratio as a function of time, showing electrosorption–desorption cycles; **b** Electrode regeneration efficiency ($\eta_r$) of $Ti_3C_2T_x$ before stabilization, $Ti_3C_2T_x$ after stabilization, and AC respectively; Error bars represent standard deviation. **c** Electrode adsorption capacity ($\Gamma$) at each electrosorption–desorption cycle for 10 wt% AC (bottom, circles) and $Ti_3C_2T_x$ (top, triangles); **d** Adsorption rates ($Q_{rate}$) for 10 wt% AC and $Ti_3C_2T_x$ flow electrodes at each electrosorption–desorption cycle.

Adsorption capacity ($\Gamma$) is an integral criterion to evaluate the electrode performance in an FE-CDI cell. The average adsorption capacities exhibit great disparity between the two electrode systems. The $\Gamma$ for $Ti_3C_2T_x$ is 460 ± 12 mg g$^{-1}$, which is more than 2 orders of magnitude higher than AC, which has an average $\Gamma$ of 4.20 ± 0.07 mg g$^{-1}$. The values for each charge–discharge cycle are shown in Fig. 4c. It can clearly be seen that $\Gamma$ decreases after the first cycle but then eventually stabilizes and remains nearly constant in the subsequent cycles. However, for $Ti_3C_2T_x$ the lowest $\Gamma$ (10th cycle) can still achieve a value of 440 mg g$^{-1}$, suggesting excellent regeneration stability (Fig. 4b). In aqueous environments, the solvated ammonium ($NH_4^+$) ions have an average radius ranging from 2.50 to 3.20 Å[95–97], which is larger than the initial interlayer spacing (2.70 Å) of MXene sheets. However, due to the hydrophilic nature of both the $Ti_3C_2T_x$ and ammonia, intercalation will readily and spontaneously occur[98]. Furthermore, with each cycle, the preintercalated lithium ions will be released, leading to more active sites on the $Ti_3C_2T_x$ basal plane, accounting for the increasing ammonia adsorption rate (Fig. 4d). The significantly higher value of ammonia $\Gamma$ follows the trend of previous studies[41,99], where 2D materials, particularly graphene, show enhanced adsorption for ammonia compared to sodium chloride (NaCl) desalination. Graphene and graphene oxide (GO) possess similar structural features as MXenes but they lack surface functional groups and the negative surface charge present in MXenes. It has been established that ammonia interacts via a combination of ion intercalation, physisorption, and chemisorption[100]. The presence of the –OH and –O functional groups on the $Ti_3C_2T_x$ surface facilitate surface reactions with the $NH_4^+$ ions. The surface chemistry and functional groups affect reactive adsorption as well as physical adsorption mechanisms[100]. Higher initial ionic concentrations, such as the 500 mg L$^{-1}$ feed solution used in this study, have been shown to enhance $\Gamma$[101]. First principle calculations on adsorption behaviors have revealed that $NH_3$ has a very small (more negative) adsorption energy ($E_{ads}$) of −0.078 eV atom$^{-1}$, which results in strong interactions with $Ti_3C_2T_x$ MXene[102]. Sensor studies have revealed that of all the considered compounds, ammonia is the only one to interact via chemisorption on the surface of MXene[103]. The calculations also show high charge transfer ($C_t$ −0.153e) between $NH_3$ and $Ti_3C_2T_x$, hence lending credibility to the hypothesis that the high $\Gamma$ value is a consequence of multiple charge transfer mechanisms[102].

The adsorption rate ($Q_{rate}$) values for the run are shown in Fig. 4d. The average adsorption rate for AC is lower (0.0021 ± 0.0003 mg min$^{-1}$ cm$^{-2}$) than $Ti_3C_2T_x$ (0.0054 ± 0.0090 mg min$^{-1}$ cm$^{-2}$) owing to the lower plateau time and higher deionization efficiency of the latter. It is interesting to note that the adsorption rate for AC changes very little across cycles, but it shows an upward trend for $Ti_3C_2T_x$ (not including the first cycle). This is a consequence of decreasing plateau times for $Ti_3C_2T_x$ electrodes. In batch system operation, $C_o$ for each subsequent cycle is different. The value of $C_o$ affects the kinetic accessibility of the dissolved ions, and hence has an effect on the adsorption rate. As discussed earlier, $\eta_r$ of AC electrodes is marginally higher than $Ti_3C_2T_x$ electrodes (Fig. 4b). This manifests as consistent $C_o$ and resultant adsorption rate for AC.

The change in total charge of the system agrees with the change in the conductivity of the effluent solution. As shown in Fig. 5a, the cell current decreases and total charge increases as ions are removed from the feed solution. During the discharging step, the current gradually rises back to its initial value, as partial charge is recovered. The current response in the system was observed at 30 mA (Fig. 5a) and is dependent upon various factors including feed solution concentration, electrical conductivity of the electrodes, contact area, and flow rates[86]. Diffusive flux between the feed solution and the flowing electrodes is shown to have a negligible effect on the current response of the system. (0.7 mA, Supplementary Fig. 1). High response current reduces overlap effect and causes an increase in the rate of ion transfer, which positively impacts the capacitance behavior and $\Gamma$[104]. The charge efficiency ($\wedge$) and total Coulombic loss ($\eta_{coul}$) in the system (Supplementary Eqs. (6) and (7)) are shown in Fig. 5b. The values for $\wedge$ range from 58% to 70% over the course of the CDI test, while roughly increasing with each subsequent cycle. This is consistent with the observed trend of increasing adsorption rate for each cycle (Fig. 4d), as $\wedge$ varies with varying $C_o$[105]. The reported values align well with results reported in literature[106]. A $\wedge$ value of 100% has never been reported. It has been theorized that the relatively low $\wedge$ values are an inherent consequence of pseudocapacitive behavior because of the presence of co-ion repulsion and counter-ion adsorption[106]. Some efforts have been made to increase the charge efficiency in CDI systems[107]. However, further work is needed in the area. The charge recovered during the discharging step is lower than the charge transferred during the charging step, resulting in a $\eta_{coul}$ increasing from 8% to 21% (Supplementary Eq. (7), Fig. 5b). This is a consequence of leakage current and is typical for supercapacitors and CDI systems. It should also be noted that current increases with each cycle (Fig. 5a), which leads to more pronounced electrode polarization and hence results in an increase in $\eta_{coul}$ with each cycle[54]. Pronounced electrode polarization is also responsible for decreasing $\Gamma$ (Fig. 4c)[54]. Barring the first cycle, the energy consumption and recovery trends (Fig. 5c) agree with $\eta_{coul}$ and Λ. This is in line with theoretical studies[107]. Energy recovery ratio is defined as the ratio of energy recovered during the release of the ions stored in the electrical double layer of the electrodes to the energy consumed during ion charge (Supplementary Eq. (10)). Over the course of ten cycles, 8–21% of energy was not recovered by the system. The profile closely resembles $\eta_{coul}$. The average energy consumption for the system was $0.45 \pm 0.01$ kWh kg$^{-1}$. This is higher than the 0.24 kWh kg$^{-1}$ obtained by Ma et al.[54], using $Ti_3C_2T_x$ films in flow-by CDI for NaCl removal. However, the adsorption capacity achieved in this work is higher, compensating for the marginally higher energy consumption. For comparison, commercial wastewater treatment plants require 4.60 kWh kg$^{-1}$ of energy for ammonium ion removal[108]. It is imperative to note that in commercial treatment plants the contaminant concentration varies widely depending on the source of wastewater. Despite the discrepancy in comparison, FE-CDI requires 10 times less energy, hence alluding to the possibility of FE-CDI being a more energy-efficient alternative.

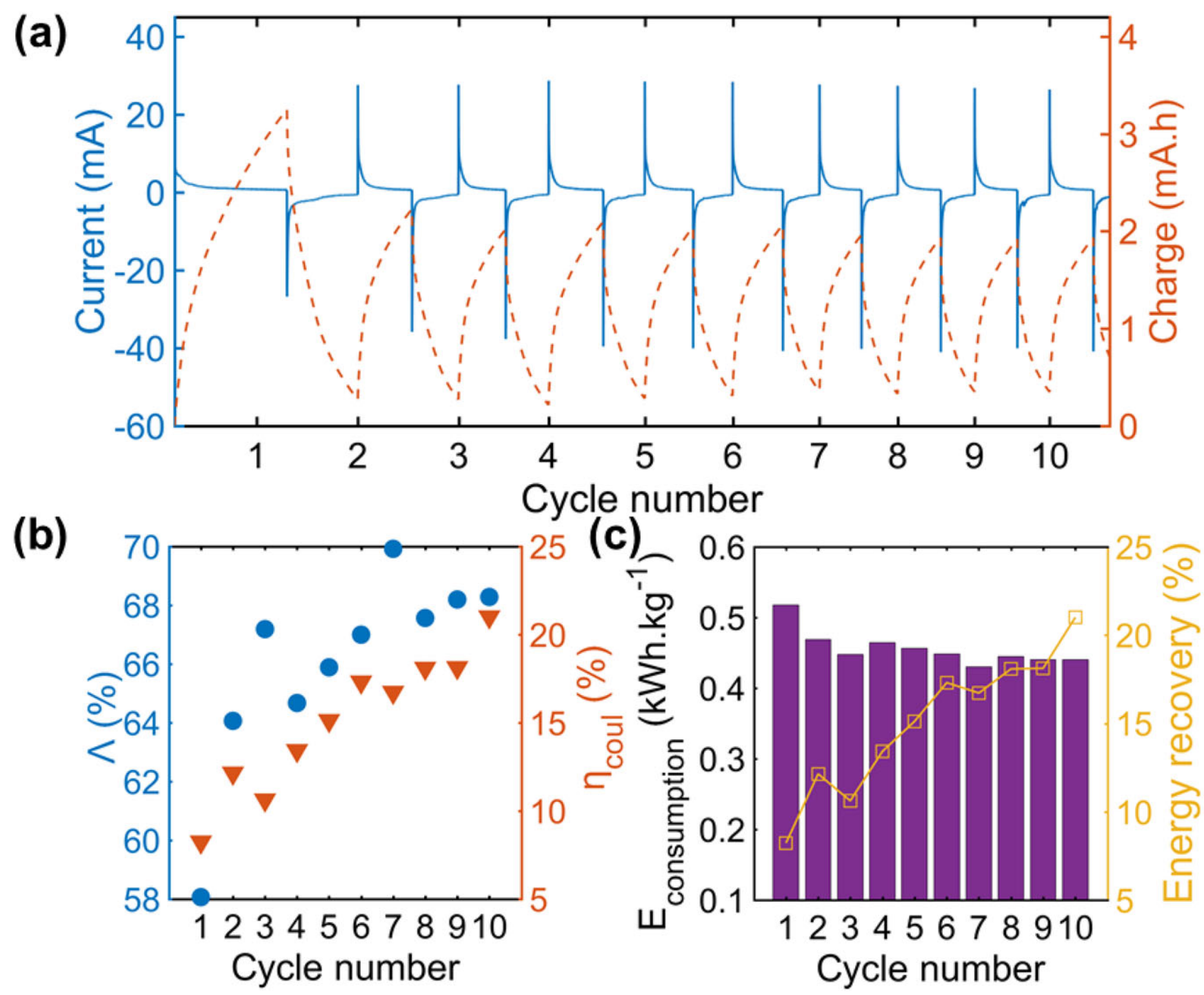

**Fig. 5 Relative energy analysis of the FE-CDI system.** For each electrosorption–desorption cycle using $Ti_3C_2T_x$ flow electrodes: **a** variation in current on left axis (solid line) and charge on right axis (dashed line); **b** charge efficiency (Λ) on left axis (circles) and coulumbic efficiency loss ($\eta_{coul}$) on right axis (triangles); **c** Energy consumption ($E_{consumption}$) on left axis (bars) and energy recovery on right axis (line and symbols).

Structure, properties, and processing correlations influence how a material will perform in a system. When comparing the structural and electrochemical properties of $Ti_3C_2T_x$ with other materials previously used for FE-CDI systems (Supplementary Table 1), it can clearly be seen that its properties allow $Ti_3C_2T_x$ to achieve high performance when employed as flow electrodes. As shown in Table 1, materials such as activated carbon, graphite, and graphene have been studied for ammonia removal in FE-CDI systems[40,41,109,110]. The adsorption capacity of $Ti_3C_2T_x$ MXene shows several orders of magnitudes improvement over 1.5 wt% graphite[99]. It is apparent that $Ti_3C_2T_x$ show markedly higher performance for ammonia removal when compared with previously researched electrode materials and CDI systems.

### Mechanisms of FE-CDI electrode performance

The capacitive charge storage mechanisms have a direct correlation with the adsorptive performance of the electrodes[111]. Figure 6a shows cyclic voltammograms (CVs) for the $Ti_3C_2T_x$ electrodes in 1 M $NH_4Cl$ aqueous electrolyte at periodically increasing scan rates. The CVs exhibit the classic quasi-rectangular shape characteristic of MXenes and other pseudocapacitive materials[112]. A broad redox peak is observed in the range of −0.35 to −0.45 V at lower scan rates of 5 and 10 mV s$^{-1}$. This is indicative of the fast and reversible redox reactions occurring at the electrode surface and is a typical characteristic of pseudocapacitive behavior[113]. In the case of layered nanomaterials like MXenes, the broad redox peaks and low peak-to-peak separation can also be indicative of intercalation because of the shortening of the ion diffusion length and all reactions being surface charge transfer dependent as opposed to solid-state diffusion, due to increased specific surface area[114]. The broad redox dip disappears at higher scan rates owing to limited diffusion and contact time. The peak broadening and flattening is a result of fast scan rate and slow kinetics of the reaction[115]. The CV curves retain their quasi-rectangular shape even at increasing scan rates, indicating the high-rate, reversible performance of $Ti_3C_2T_x$ electrodes. This agrees with the repeatable behavior shown during deionization testing (Fig. 4a), and is further substantiated by the almost linear galvanostatic charge–discharge (GCD) profiles shown in Fig. 6b. The near linear nature of the GCD curves indicates rapid and reversible charge and discharge in the electrical double layer[116]. The double layer storage effects can also be observed in CVs (Fig. 6a) as the ends of the forward and reverse scans taper off (informally referred to as "curves taking off")[117].

**Table 1.** . Comparison of adsorption capacity and deionization efficiency between different electrode materials and operational modes of capacitive deionization.

| Electrode material | Ion species | Cell architecture | Applied voltage (V) | Current | Initial concentration (mg.L$^{-1}$) | Deionization efficiency, $\eta_r$ (%) | Adsorption capacity, $\Gamma$ (mg.g$^{-1}$) | Reference |
|---|---|---|---|---|---|---|---|---|
| $K_2Ti_2O_5$-Activated Carbon | $NH_4^+$ | Flow electrode | 1.2 | - | 53.5 | 65 | - | 40 |
| Graphene Laminates | $NH_4Cl$ | Membrane assisted | 2.0 | 0.17 A | 400 | 99 | 15.30 | 41 |
| $Ti_3C_2T_x$ MXene | NaCl | Flow by | 1.2 | Constant current – 20 mA | 585 | - | 68 | 54 |
| Porous $Ti_3C_2T_x$ MXene | NaCl | Flow by | 1.2 | - | 10000 | - | 45 | 55 |
| $Ti_3C_2T_x$ MXene | NaCl | Flow by | 1.2 | - | 0.085 | - | 13 | 56 |
| Ar-$Ti_3C_2T_x$ MXene | NaCl | Flow by | 0.8–1.6 | - | 500 | - | 26.80 | 57 |
| Graphite (1.5 wt%) | $NH_4Cl$ | Flow electrode | 0.2–1.2 | - | 20 | 87 | 1.43 | 85 |
| Carbon Cloth | $NH_4^+$–N | Flow by | 1.2–3.0 | - | 68.8 | 60.5–95.7 | - | 109 |
| Vertically Oriented Graphenes | NaCl | Flow through | 1.0 | - | 5000 | 98.1 | 165 | 110 |
| **$Ti_3C_2T_x$ MXene** | **$NH_4Cl$** | **Flow electrode** | **1.2** | **30 mA** | **500** | **60** | **460** | **This Work** |

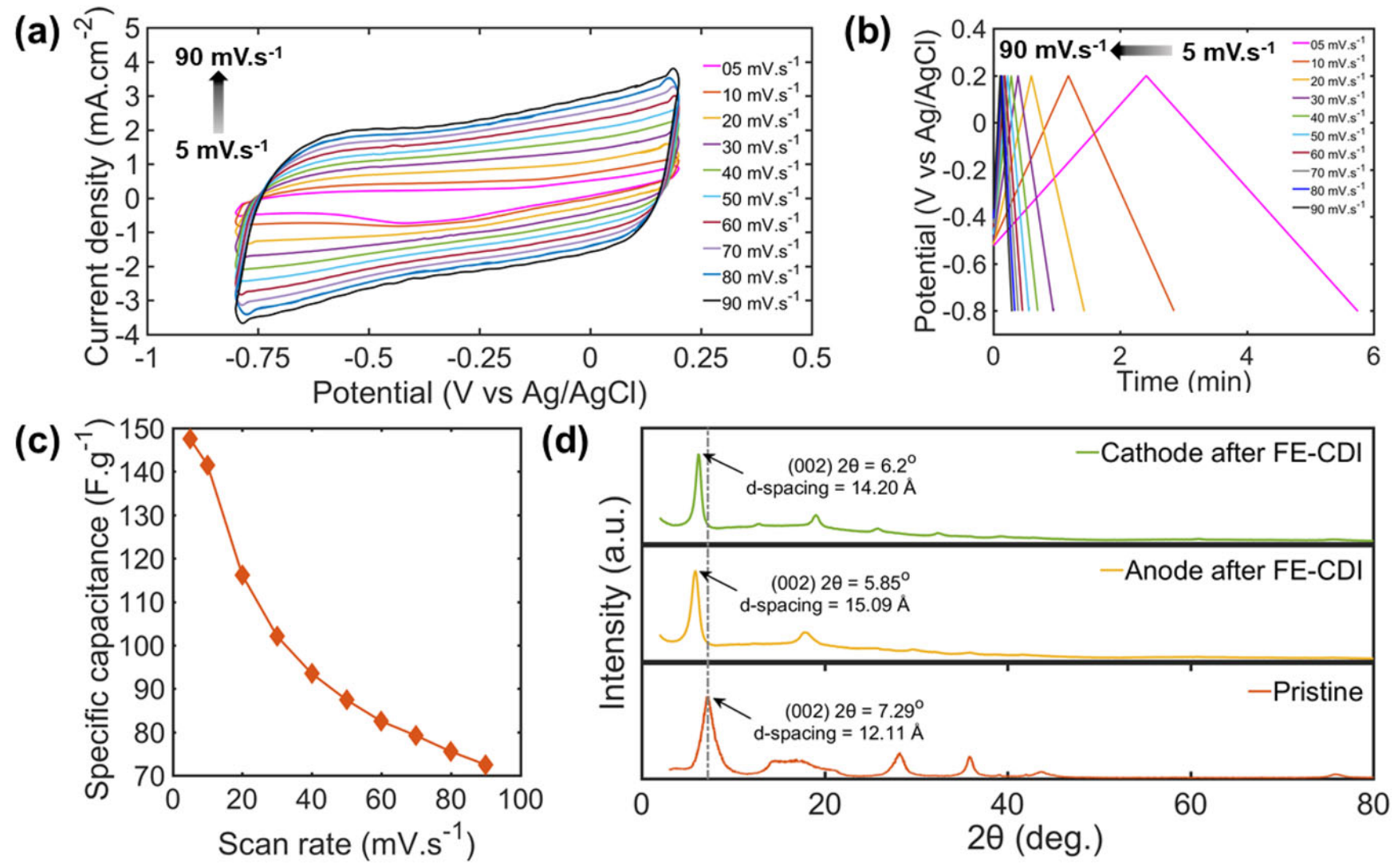

**Fig. 6 Post-mortem analysis of FE-CDI electrodes.** For $Ti_3C_2T_x$ electrode in 1M $NH_4Cl$: **a** cyclic voltammograms at varying scan rates; **b** galvanostatic charge/discharge profiles at varying scan rates; **c** specific capacitance as a function of scan rate; and **d** XRD patterns before and after FE-CDI operation.

The changes in specific capacitance as a function of varying scan rates are depicted in Fig. 6c. $Ti_3C_2T_x$ electrode in $NH_4Cl$ aqueous electrolyte delivers a high specific capacitance of 148 F g$^{-1}$ at 5 mV s$^{-1}$. The total capacitance in pseudocapacitive materials is a combination of redox (Faradaic and non-Faradaic) reactions and intercalation pseudocapacitance[114]. Owing to combined effect, per atom charge storage is greater than for electrical double layer based capacitors[114]. Lower scan rates exhibited higher $C_p$ values with continuous decrease reaching the minimum of 73 F g$^{-1}$ at 90 mV s$^{-1}$. Ion adsorption in CDI is a time intensive and flow-rate-dependent process. Hence, higher scan rates do not reflect all the mechanisms at play. This is because the electrolyte ions can easily access the electrode surface at lower scan rates resulting in significant enhancement in the electrosorption capacitance[113]. The reported specific capacitance values are in agreement with previous studies that show that $Ti_3C_2T_x$ electrodes exhibit maximum charge storage capacities in the presence of polar protic electrolytes[74].

Figure 6d shows the post-mortem XRD pattern of the $Ti_3C_2T_x$ positive and negative electrodes before and after the partial charge cycle. No significant difference in the diffraction intensity of the peaks was observed (Supplementary Fig. 3). Using Bragg's law, the position of the (002) peak is used to calculate the *d*-spacing of the MXene crystal structure[67]. Typically, dry $Ti_3C_2T_x$ exhibits a (002) peak at 9.40° (*d*-spacing of 9.41 Å); for the pristine $Ti_3C_2T_x$ in this study, the (002) peak is initially at 7.29°, corresponding to a *d*-spacing of 12.11 Å, or an interlayer spacing of 2.70 Å. The primary peak shifts to a lower angle of 5.85° (*d*-spacing of 15.09 Å) for the anode. The cathode peak also shifts to a smaller $2\theta$ value of 6.20° (*d*-spacing of 14.20 Å). This corresponds to an increase in the interlayer spacing of 2.98 and 2.09 Å (total interlayer spacing of 5.68 and 4.79 Å) for the anode and cathode, respectively. The expanded crystal structure is due to ion entrapment and intercalation in between the MXene layers. The anodic side has slightly greater expansion due to the presence of like repulsive charges[57]. Based on the expanded interlayer spacing and ionic sizes, it can be estimated that on average there are ~2 layers of ions entrapped between the MXene layers. This agrees with previous studies on capacitive deionization and adsorption[57,118–120]. This supports the pseudocapacitive behavior observed in CVs (Fig. 6a), that no single charge storage mechanism is dominant and the capacitive behavior of $Ti_3C_2T_x$ electrodes is a combination of surface redox, electrical double layer, and ion intercalation phenomena.

## DISCUSSION

In this study, we demonstrated an FE-CDI system with high ammonia removal performance based on flow electrodes consisting of $Ti_3C_2T_x$ MXene. An average adsorption capacity of 460 mg g$^{-1}$ along with a low-energy consumption of 0.45 kWh kg$^{-1}$ was measured. Electrochemical analysis revealed a high specific capacitance of 148 F g$^{-1}$ in ammonia electrolyte. The increase in interlayer spacing revealed by XRD investigations show evidence for intercalation-assisted ion storage. The results reinforce the strong dependence of FE-CDI performance on the characteristics of the electrode material. Owing to its high conductivity, colloidal stability, high surface area, and unique surface chemistry $Ti_3C_2T_x$ is a promising candidate for ammonia removal and recovery from industrial and commercial wastewaters.

While the results presented are promising, further research, including theoretical modeling of the kinetic analysis, and testing different MXenes is needed to further improve the process. Due to the non-uniform surface chemistry on MXene surfaces, it is important to consider how different etching approaches will change the adsorption process. Finally, MXenes have been shown as viable for many adsorption processes, but the studies are still in their infancy, $Ti_3C_2T_x$ and other MXenes should be tested for adsorption of more pollutants. However, based on these results, $Ti_3C_2T_x$ is among the most promising environmental remediation materials for ammonia removal from wastewater systems. We expect that the demonstrated work will open new avenues for realizing high performance, energy-efficient, large-scale pollutant removal systems.

## METHODS

### Preparation of activated carbon flow electrodes

The 10 wt% control (carbon) flow electrodes were prepared by mixing 1.20 g of 80 mesh activated carbon (AC) powder (Cabot Norit® A Ultra E 153) in 12 mL of ultrapure water (18 MΩ-cm) and stirring for 2 h. The mixture was probe sonicated for 1 h at 55 W to reduce the particle size by breaking up agglomerates, increasing the flowability. During cell operation, the electrodes were continuously stirred in the electrode solution containers, using a magnetic stir bar to prevent sedimentation of the carbon particles.

### Preparation of $Ti_3C_2T_x$ flow electrodes

$Ti_3C_2T_x$ MXene was etched from $Ti_3AlC_2$ MAX phase powder using the minimally intensive layer delamination (MILD) synthesis method[121]. This method was selected because of its reduced toxicity, and the ability to produce low-defect, larger MXene flakes[122,123]. The fluoride containing wet etching process results in –O, –OH, and –F surface terminations which contribute to the hydrophilic nature of the MXene flakes[81]. 20 mL of 9 M hydrochloric acid (HCl, Alfa Aesar) was stirred with 1.60 g of lithium fluoride (LiF, 99.85% Alfa Aesar) using a Teflon coated magnetic bar at 300 rpm for 10 min prior to addition of the MAX. 1 g of $Ti_3AlC_2$ was added to the in-situ synthesized HF solution in four increments to prevent overheating of the solution. The reaction was allowed to run for 24 h at room temperature and ambient pressure (RTP). The resultant mixture was then washed using ultrapure water (18 MΩ-cm) via centrifugation at 3500 rpm until the acidic supernatant became neutral (pH 5–6). The presence of lithium ions ($Li^+$) with the etching solution causes simultaneous etching/delamination resulting in an electrostatically stable colloidal solution. The stable supernatant was vacuum filtered with a glass microfiber filter (0.45 µm, Whatman) to separate the MXene clay. To prepare the flow electrode solution, the MXene residue was re-dispersed in ultrapure water (18 MΩ-cm) via manual shaking to obtain a solution with concentration of 1 mg mL$^{-1}$.

### Characterization of $Ti_3C_2T_x$

To evaluate the morphology and structure of the samples, several characterization techniques were employed. X-ray diffraction (XRD) was conducted on a Rigaku Smartlab XRD system equipped with a Cu $K_\alpha$ source, a step size of 0.03° 2$\theta$, and a holding time of 0.5 s from 3° to 90° 2$\theta$. The characteristic expanded structure of the MXenes was observed via scanning electron microscopy (SEM) and transmission electron microscopy (TEM) imaging on a FEI Teneo field emission SEM and JEOL JEM-2100 HR analytical TEM, respectively. Microscopic imaging was also used to map out lateral size of the flakes, which was further confirmed by dynamic light scattering (DLS) performed on a Brookhaven NanoBrook Omni. The obtained measurement is the hydrodynamic diameter, where the particle is assumed to be spherical in nature[124]. The measured diameter is a function of the diffusion coefficient according to the Stokes–Einstein relation[124]. The measurement is an effective average, and hence can deviate from actual lateral flake size. The tool was also used to measure the zeta ($\zeta$) potential of the colloidal suspension. All measurements were taken at RTP (25 °C, 1 atm) and at a pH of 7. The viscosity of the colloidal suspension was measured on a RheoSense µVisc viscometer with 300 µL of solution to obtain five repeated measurements. Raman spectra were obtained using a 532 nm He−Cd laser on a Horiba LabRAM HR Evolution Raman after drop casting the electrode solution on a glass slide.

### Capacitive deionization experiments

To evaluate the performance of the electrodes a self-assembled FE-CDI unit with titanium current collectors (6.35 cm × 6.35 cm) containing serpentine flow channels was operated in batch mode in constant voltage (1.20 V) cycle in both charging and discharging stages. The power was sourced from a Biologic SP-50 potentiostat. The effective transfer area was 10 cm$^2$. As shown in Fig. 1b, the cell was assembled with vitreous carbon (Duocel Reticulated Vitreous Carbon 100 ppi), carbon cloth (AvCarb 1071 HCB), rubber gaskets (Neoprene, 1.5 mm), along with pre-treated anion and cation exchange membranes (Fumasep FAA-3-PK-130 and Nafion™ 115, respectively). These were separated by a non-conductive porous spacer (Nylon 3/64″) with a polyester filter felt (50 µm) that allowed the feed water to pass through. 20 mL of 500 mg L$^{-1}$ ammonium chloride ($NH_4Cl$) solution was prepared by dissolving analytical grade $NH_4Cl$ (99.99% Sigma Aldrich) in ultrapure water (18 MΩ-cm). The solution was circulated through the cell at a flow rate of 2 mL min$^{-1}$ while a conductivity meter (HACH H1440d Benchtop Meter) monitored the conductivity of the $NH_4Cl$ solution. 6 mL each of the two electrode solutions were circulated in the cell channels at a flow rate of 3 mL min$^{-1}$.

### Electrochemical characterization

The electrochemical response of the $Ti_3C_2T_x$ electrodes was studied using a three-electrode cell setup with 1 M $NH_4Cl$ electrolyte. MXene flow electrode solution was used to prepare a freestanding film via vacuum-assisted filtration onto a polyvinylidene fluoride (PVDF) filter (0.22 µm, Durapore). To yield a freestanding film, the residue was peeled from the filter after drying in vacuum at room temperature for 24 h. 1 cm$^2$ of the freestanding film was employed as the working electrode, with a standard platinum wire counter electrode, and Ag/AgCl (SYC Technologies) in 1 M KCl (Sigma Aldrich, 99%) as the reference electrode. The cyclic voltammetry (CV), using scan rates of 5 to 90 mV s$^{-1}$, and galvanostatic charge–discharge (GCD) tests were conducted using the Biologic SP-50 potentiostat.

## DATA AVAILABILITY

The data used to support the findings of this study is included within the article (and its supplementary information files).

## REFERENCES

1. Macknick, J., Newmark, R., Heath, G. & Hallett, K. C. Operational water consumption and withdrawal factors for electricity generating technologies: a review of existing literature. *Environ. Res. Lett.* **7**, 045802 (2012).
2. Dept. of Energy. *Energy Demands on Water Resources: Report to Congress on the Interdependency of Energy and Water* (U.S. Department of Energy, 2006).
3. Intl. Energy Agency. *World Energy Outlook* (OECD Publishing, International Energy Agency, 2019).
4. Palo, M., Uusivuori, J., Mery, G., Korotkov, A. V. & Humphreys, D. *World Forests, Markets and Policies: Towards a Balance* (Springer, Dordrecht, 2001).
5. USGS. *Mineral Commodity Summaries* (U.S. Geological Survey, 2020).
6. U.S. EPA. *Ambient Water Quality Criteria for Ammonia* (U.S. Environmental Protection Agency, 2013).
7. Visek, W. J. Ammonia: its effects on biological systems, metabolic hormones, and reproduction. *J. Dairy Sci.* **67**, 481–498 (1984).
8. Gruber, N. & Galloway, J. N. An earth-system perspective of the global nitrogen cycle. *Nature* **451**, 293–296 (2008).
9. Mao, G. et al. Technology status and trends of industrial wastewater treatment: a patent analysis. *Chemosphere* **288**, 132483 (2022).
10. Wang, Z., Zhao, X., Xie, T., Wen, N. & Yao, J. A Comprehensive evaluation model of ammonia pollution trends in a groundwater source area along a river in residential areas. *Water* **13**, 1924 (2021).
11. Chai, W. S., Bao, Y., Jin, P., Tang, G. & Zhou, L. A review on ammonia, ammonia–hydrogen and ammonia–methane fuels. *Renew. Sustain. Energy Rev*. **147**, 111254 (2021).
12. De La Hoz, R. E., Schlueter, D. P. & Rom, W. N. Chronic lung disease secondary to ammonia inhalation injury: a report on three cases. *Am. J. Ind. Med.* **29**, 209–214 (1996).
13. Keiser, David A.J.S.S. Consequences of the Clean Water Act and the demand for water quality. *Q. J. Econ.* **134**, 349–396 (2019).
14. Wu, J., Cao, M., Tong, D., Finkelstein, Z. & Hoek, E. M. V. A Critical review of point-of-use drinking water treatment in the United States. *npj Clean Water* **4**, 40 (2021).
15. Karri, R. R., Sahu, J. N. & Chimmiri, V. Critical review of abatement of ammonia from wastewater. *J. Mol. Liq.* **261**, 21–31 (2018).
16. Han, B., Butterly, C., Zhang, W., He, J. & Chen, D. Adsorbent materials for ammonium and ammonia removal: a review. *J. Clean. Prod*. **283**, 124611 (2021).
17. Rohani, R. et al. Ammonia removal from raw water by using adsorptive membrane filtration process. *Sep. Purif. Technol*. **270**, 118757 (2021).
18. Eskicioglu, C., Galvagno, G. & Cimon, C. Approaches and processes for ammonia removal from side-streams of municipal effluent treatment plants. *Bioresour. Technol.* **268**, 797–810 (2018).
19. Adam, M. R. et al. Current Trends and future prospects of ammonia removal in wastewater: a comprehensive review on adsorptive membrane development. *Sep. Purif. Technol.* **213**, 114–132 (2019).
20. Manasa, R. L. & Mehta, A. Current perspectives of anoxic ammonia removal and blending of partial nitrifying and denitrifying bacteria for ammonia reduction in wastewater treatment. *J. Water Process Eng*. **41**, 102085 (2021).
21. Jaramillo, F., Orchard, M., Muñoz, C., Zamorano, M. & Antileo, C. Advanced strategies to improve nitrification process in sequencing batch reactors—a review. *J. Environ. Manag.* **218**, 154–164 (2018).
22. Vineyard, D., Hicks, A., Karthikeyan, K. G. & Barak, P. Economic analysis of electrodialysis, denitrification, and anammox for nitrogen removal in municipal wastewater treatment. *J. Clean. Prod*. **262**, 121145 (2020).
23. Yang, J. & Chen, B. Energy Efficiency Evaluation of Wastewater Treatment Plants (WWTPs) Based on Data Envelopment Analysis. *Appl. Energy* **289**, 116680 (2021).
24. Rouwenhorst, K. H. R., Van der Ham, A. G. J., Mul, G. & Kersten, S. R. A. Islanded ammonia power systems: technology review & conceptual process design. *Renew. Sustain. Energy Rev*. **114**, 109339 (2019).
25. Pfromm, P. H. Towards sustainable agriculture: fossil-free ammonia. *J. Renew. Sustain. Energy* **9**, 034702 (2017).
26. Blair, J. W. & Murphy, G. W. Electrochemical demineralization of water with porous electrodes of large surface area. In *Saline Water Conversion* (eds American Chemical Society) 206–223 (American Chemical Society, 1960).
27. Anderson, M. A., Cudero, A. L. & Palma, J. Capacitive deionization as an electrochemical means of saving energy and delivering clean water. comparison to present desalination practices: will it compete? *Electrochim. Acta* **55**, 3845–3856 (2010).
28. Ahmed, M. A. & Tewari, S. Capacitive deionization: processes, materials and state of the technology. *J. Electroanal. Chem.* **813**, 178–192 (2018).
29. Tang, K., Kim, Y.-H., Yiacoumi, S. & Tsouris, C. (Invited) Capacitive deionization of high-salinity water using ion-exchange membranes. *ECS Meet. Abstr.* **MA2017-02**, 1032 (2017).
30. AlMarzooqi, F. A., Al Ghaferi, A. A., Saadat, I. & Hilal, N. Application of capacitive deionisation in water desalination: a review. *Desalination* **342**, 3–15 (2014).
31. Oren, Y. Capacitive deionization (CDI) for desalination and water treatment—past, present and future (a review). *Desalination* **228**, 10–29 (2008).
32. Tang, W. et al. Various cell architectures of capacitive deionization: recent advances and future trends. *Water Res.* **150**, 225–251 (2019).
33. Presser, V. et al. The electrochemical flow capacitor: a new concept for rapid energy storage and recovery. *Adv. Energy Mater.* **2**, 895–902 (2012).
34. Hatzell, K. B. et al. Capacitive deionization concept based on suspension electrodes without ion exchange membranes. *Electrochem. Commun.* **43**, 18–21 (2014).
35. Biesheuvel, P. M., Van Limpt, B. & Van der Wal, A. Dynamic adsorption/desorption process model for capacitive deionization. *J. Phys. Chem. C* **113**, 5636–5640 (2009).
36. Remillard, E. M., Shocron, A. N., Rahill, J., Suss, M. E. & Vecitis, C. D. A direct comparison of flow-by and flow-through capacitive deionization. *Desalination* **444**, 169–177 (2018).
37. Dahiya, S. & Mishra, B. K. Enhancing understandability and performance of flow electrode capacitive deionisation by optimizing configurational and operational parameters: a review on recent progress. *Sep. Purif. Technol.* **240**, 116660 (2020).
38. Yang, F. et al. Flow-electrode capacitive deionization: a review and new perspectives. *Water Res*. **200**, 117222 (2021).
39. Ratajczak, P., Suss, M. E., Kaasik, F. & Béguin, F. Carbon electrodes for capacitive technologies. *Energy Storage Mater.* **16**, 126–145 (2019).
40. Lin, L. et al. Selective ammonium removal from synthetic wastewater by flow-electrode capacitive deionization using a novel $K_2Ti_2O_5$-activated carbon mixture electrode. *Environ. Sci. Technol.* **54**, 12723–12731 (2020).
41. Wimalasiri, Y., Mossad, M. & Zou, L. Thermodynamics and kinetics of adsorption of ammonium ions by graphene laminate electrodes in capacitive deionization. *Desalination* **357**, 178–188 (2015).
42. Luciano, M. A., Ribeiro, H., Bruch, G. E. & Silva, G. G. Efficiency of capacitive deionization using carbon materials based electrodes for water desalination. *J. Electroanal. Chem*. **859**, 113840 (2020).
43. Choo, K. Y., Yoo, C. Y., Han, M. H. & Kim, D. K. Electrochemical analysis of slurry electrodes for flow-electrode capacitive deionization. *J. Electroanal. Chem.* **806**, 50–60 (2017).
44. Shin, Y.-U., Lim, J., Boo, C. & Hong, S. Improving the feasibility and applicability of flow-electrode capacitive deionization (FCDI): review of process optimization and energy efficiency. *Desalination* **502**, 114930 (2021).
45. Porada, S. et al. Carbon flow rlectrodes for continuous operation of capacitive deionization and capacitive mixing energy generation. *J. Mater. Chem. A* **2**, 9313–9321 (2014).
46. Deysher, G. et al. Synthesis of $Mo_4VAlC_4$ MAX phase and two-dimensional $Mo_4VC_4$ MXene with five atomic layers of transition metals. *ACS Nano* **14**, 204–217 (2020).
47. Shuck, C. E. et al. Scalable synthesis of $Ti_3C_2T_x$ MXene. *Adv. Eng. Mater*. **22**, 1901241 (2020).
48. Vahid Mohammadi, A., Rosen, J. & Gogotsi, Y. The world of two-dimensional carbides and nitrides (MXenes). *Science*. **372**, 6547 (2021).
49. Zhang, Y., Wang, L., Zhang, N. & Zhou, Z. Adsorptive environmental applications of MXene nanomaterials: a review. *RSC Adv.* **8**, 19895–19905 (2018).
50. Liu, G. et al. Ultrathin two-dimensional MXene membrane for pervaporation desalination. *J. Membr. Sci.* **548**, 548–558 (2018).
51. Rasool, K. et al. Water treatment and environmental remediation applications of two-dimensional metal carbides (MXenes). *Mater. Today* **30**, 80–102 (2019).
52. Naguib, M. et al. Two-dimensional nanocrystals: two-dimensional nanocrystals produced by exfoliation of $Ti_3AlC_2$. *Adv. Mater.* **23**, 4248–4253 (2011).
53. Nasrallah, G. K., Al-Asmakh, M., Rasool, K. & Mahmoud, K. A. Ecotoxicological assessment of $Ti_3C_2T_X$ (MXene) using a Zebrafish Embryo Model. *Environ. Sci. Nano* **5**, 1002–2022 (2018).
54. Ma, J., Cheng, Y., Wang, L., Dai, X. & Yu, F. Free-standing $Ti_3C_2T_x$ MXene film as binder-free electrode in capacitive deionization with an ultrahigh desalination capacity. *Chem. Eng. J.* **384**, 123329 (2020).
55. Bao, W. et al. Porous cryo-dried MXene for efficient capacitive deionization. *Joule* **2**, 778–787 (2018).
56. Agartan, L. et al. Influence of operating conditions on the desalination performance of a symmetric pre-conditioned $Ti_3C_2T_x$-MXene membrane capacitive deionization system. *Desalination* **477**, 114267 (2020).
57. Srimuk, P. et al. MXene as a novel intercalation-type pseudocapacitive cathode and anode for capacitive deionization. *J. Mater. Chem. A* **4**, 18265–18271 (2016).
58. Guo, L. et al. Ar plasma modification of 2D MXene $Ti_3C_2T_x$ nanosheets for efficient capacitive desalination. *FlatChem* **8**, 17–24 (2018).

59. Venkateshalu, S. & Grace, A. N. MXenes—a new class of 2D layered materials: synthesis, properties, applications as supercapacitor electrode and beyond. *Appl. Mater. Today* **18**, 100509 (2020).
60. Anasori, B., Lukatskaya, M. R. & Gogotsi, Y. 2D Metal carbides and nitrides (MXenes) for energy storage. *Nat. Rev. Mater*. **2**, 16098 (2017).
61. Peng, Y. Y. et al. All-MXene (2D titanium carbide) solid-state micro-supercapacitors for on-chip energy storage. *Energy Environ. Sci.* **9**, 2847–2854 (2016).
62. Zhang, X., Zhang, Z. & Zhou, Z. MXene-based materials for electrochemical energy storage. *J. Energy Chem.* **27**, 73–85 (2018).
63. Shukla, A. K., Banerjee, A., Ravikumar, M. K. & Jalajakshi, A. Electrochemical capacitors: technical challenges and prognosis for future markets. *Electrochim. Acta* **84**, 165–173 (2012).
64. Simon, P. & Gogotsi, Y. Materials for electrochemical capacitors. *Nat. Mater.* **7**, 845–854 (2008).
65. Zhao, Y. et al. Performance comparison and energy consumption analysis of capacitive deionization and membrane capacitive deionization processes. *Desalination* **324**, 127–133 (2013).
66. Gogotsi, Y. & Anasori, B. The rise of MXenes. *ACS Nano* **13**, 8491–8494 (2019).
67. Shekhirev, M., Shuck, C. E., Sarycheva, A. & Gogotsi, Y. Characterization of MXenes at every step, from their precursors to single flakes and assembled films. *Prog. Mater. Sci*. **120**, 100757 (2021).
68. Singh, K., Porada, S., de Gier, H. D., Biesheuvel, P. M. & de Smet, L. C. P. M. Timeline on the application of intercalation materials in capacitive deionization. *Desalination* **455**, 115–134 (2019).
69. Lu, M., Han, W., Li, H., Zhang, W. & Zhang, B. There is plenty of space in the MXene layers: the confinement and fillings. *J. Energy Chem.* **48**, 344–363 (2020).
70. Sarycheva, A. & Gogotsi, Y. Raman spectroscopy analysis of the structure and surface chemistry of $Ti_3C_2T_x$ MXene. *Chem. Mater.* **32**, 3480–3488 (2020).
71. Cuesta, A., Dhamelincourt, P., Laureyns, J., Martínez-Alonso, A. & Tascón, J. M. D. Raman microprobe studies on carbon materials. *Carbon N. Y.* **32**, 1523–1532 (1994).
72. Schwan, J., Ulrich, S., Batori, V., Ehrhardt, H. & Silva, S. R. P. Raman spectroscopy on amorphous carbon films. *J. Appl. Phys.* **80**, 440–447 (1996).
73. Daud, W. M. A. W. & Houshamnd, A. H. Textural characteristics, surface chemistry and oxidation of activated carbon. *J. Nat. Gas Chem.* **19**, 267–279 (2010).
74. Lukatskaya, M. R. et al. Ultra-high-rate pseudocapacitive energy storage in two-dimensional transition metal carbides. *Nat. Energy* **2**, 17105 (2017).
75. Pecora, R. Doppler shifts in light scattering from pure liquids and polymer solutions. *J. Chem. Phys.* **40**, 1604–1614 (1964).
76. Zimm, B. H. Apparatus and methods for measurement and interpretation of the angular variation of light scattering; preliminary results on polystyrene solutions. *J. Chem. Phys.* **16**, 1099–1116 (1948).
77. Litan, A. Fluctuation theory of light scattering from liquids. *J. Chem. Phys.* **48**, 1052–1058 (1968).
78. Hershey, A. D., Burgi, E. & Ingraham, L. Sedimentation coefficient and fragility under hydrodynamic shear as measures of molecular weight of the DNA of phage T5. *Biophys. J.* **2**, 423–431 (1962).
79. Maleski, K., Mochalin, V. N. & Gogotsi, Y. Dispersions of two-dimensional titanium carbide MXene in organic solvents. *Chem. Mater.* **29**, 1632–1640 (2017).
80. Halim, J. et al. X-Ray photoelectron spectroscopy of select multi-layered transition metal carbides (MXenes). *Appl. Surf. Sci.* **362**, 406–417 (2016).
81. Anayee, M. et al. Role of acid mixtures etching on the surface chemistry and sodium ion storage in $Ti_3C_2T_x$ MXene. *Chem. Commun.* **56**, 6090–6093 (2020).
82. Rahaman, M. N. Science of colloidal processing. In *Ceramic Processing* (eds Taylor & Francis Group) 141–189 (CRC Press, 2017).
83. Samimi, S., Maghsoudnia, N., Eftekhari, R. B. & Dorkoosh, F. Lipid-based nanoparticles for drug delivery systems. In *Characterization and Biology of Nanomaterials for Drug Delivery* (eds Mohapatra, S., Ranjan, S., Dasgupta, N., Mishra, R. & Thomas. S.) 47–76 (Elsevier, 2019).
84. Nakatuka, Y., Yoshida, H., Fukui, K. & Matuzawa, M. The effect of particle size distribution on effective zeta-potential by use of the sedimentation method. *Adv. Powder Technol.* **26**, 650–656 (2015).
85. Fang, K. et al. Recovering ammonia from municipal wastewater by flow-electrode capacitive deionization. *Chem. Eng. J.* **348**, 301–309 (2018).
86. Porada, S., Zhao, R., Van Der Wal, A., Presser, V. & Biesheuvel, P. M. Review on the science and technology of water desalination by capacitive deionization. *Prog. Mater. Sci.* **58**, 1388–1442 (2013).
87. Aslam, M. K., Niu, Y. & Xu, M. MXenes for non-lithium-ion (Na, K, Ca, Mg, and Al) batteries and supercapacitors. *Adv. Energy Mater*. **11**, 2000681 (2021).
88. Aslam, M. K. et al. 2D MXene materials for sodium ion batteries: a review on energy storage. *J. Energy Storage* **37**, 102478 (2021).
89. Gao, Q. et al. Tracking ion intercalation into layered $Ti_3C_2$ MXene films across length scales. *Energy Environ. Sci.* **13**, 2549–2558 (2020).
90. Sang, X. et al. Atomic defects in monolayer titanium carbide ($Ti_3C_2T_x$) MXene. *ACS Nano* **10**, 9193–9200 (2016).
91. Habib, T. et al. Oxidation stability of $Ti_3C_2T_x$ MXene nanosheets in solvents and composite films. *npj 2D Mater. Appl.* **3**, 8 (2019).
92. Lee, Y. et al. Oxidation-resistant titanium carbide MXene film. *J. Mater. Chem. A* **8**, 573–581 (2019).
93. Mathis, T. S. et al. Modified MAX phase synthesis for environmentally stable and highly conductive $Ti_3C_2$ MXene. *ACS Nano* **15**, 6420–6429 (2021).
94. Natu, V. et al. 2D $Ti_3C_2T_z$ MXene synthesized by water-free etching of $Ti_3AlC_2$ in polar organic solvents. *Chem* **6**, 616–630 (2020).
95. Hollas, D. et al. Aqueous solution chemistry of ammonium cation in the auger time window. *Sci. Rep.* **7**, 756 (2017).
96. Guo, J. et al. Hydration of $NH_4^+$ in water: bifurcated hydrogen bonding structures and fast rotational dynamics. *Phys. Rev. Lett*. **125**, 106001 (2020).
97. Sidey, V. On the effective ionic radii for ammonium. *Acta Crystallogr. B. Struct. Sci. Cryst. Eng. Mater.* **72**, 626–633 (2016).
98. Chmiola, J. et al. Anomalous increase in carbon capacitance at pore sizes less than 1 nanometer. *Science* **313**, 1760–1763 (2006).
99. Seredych, M. & Bandosz, T. J. Removal of ammonia by graphite oxide via its intercalation and reactive adsorption. *Carbon N. Y.* **45**, 2130–2132 (2007).
100. Wang, S., Sun, H., Ang, H. M. & Tadé, M. O. Adsorptive remediation of environmental pollutants using novel graphene-based nanomaterials. *Chem. Eng. J.* **226**, 336–347 (2013).
101. Moreno, D. & Hatzell, M. C. Influence of feed-electrode concentration differences in flow-electrode systems for capacitive deionization. *Ind. Eng. Chem. Res.* **57**, 8802–8809 (2018).
102. Wu, M. et al. $Ti_3C_2$ MXene-based sensors with high selectivity for $NH_3$ detection at room temperature. *ACS Sens.* **4**, 2763–2770 (2019).
103. Mehdi Aghaei, S., Aasi, A. & Panchapakesan, B. Experimental and theoretical advances in MXene-based gas sensors. *ACS Omega* **6**, 2450–2461 (2021).
104. Li, Y. et al. A protic salt-derived porous carbon for efficient capacitive deionization: balance between porous structure and chemical composition. *Carbon N. Y.* **116**, 21–32 (2017).
105. Biesheuvel, P. M., Porada, S., Levi, M. & Bazant, M. Z. Attractive forces in microporous carbon electrodes for capacitive deionization. *J. Solid State Electrochem.* **18**, 1365–1376 (2014).
106. Kim, T. et al. Enhanced charge efficiency and reduced energy use in capacitive deionization by increasing the discharge voltage. *J. Colloid Interface Sci.* **446**, 317–326 (2015).
107. Gao, X., Omosebi, A., Landon, J. & Liu, K. Enhancement of charge efficiency for a capacitive deionization cell using carbon xerogel with modified potential of zero charge. *Electrochem. Commun.* **39**, 22–25 (2014).
108. Ekman, M., Björlenius, B. & Andersson, M. Control of the aeration volume in an activated sludge process using supervisory control strategies. *Water Res.* **40**, 1668–1676 (2006).
109. Ge, Z., Chen, X., Huang, X. & Ren, Z. J. Capacitive deionization for nutrient recovery from wastewater with disinfection capability. *Environ. Sci. Water Res. Technol.* **4**, 33–39 (2018).
110. Bo, Z. et al. Photo-electric capacitive deionization enabled by solar-driven nanoionics on the edges of plasma-made vertical graphenes. *Chem. Eng. J.* **422**, 130156 (2021).
111. Fic, K. et al. Revisited insights into charge storage mechanisms in electrochemical capacitors with $Li_2SO_4$-based electrolyte. *Energy Storage Mater.* **22**, 1–14 (2019).
112. Zhan, C. et al. Understanding the MXene pseudocapacitance. *J. Phys. Chem. Lett.* **9**, 1223–1228 (2018).
113. Jiang, Y. & Liu, J. Definitions of pseudocapacitive materials: a brief review. *Energy Environ. Mater.* **2**, 30–37 (2019).
114. Yu, X. et al. Emergent pseudocapacitance of 2D nanomaterials. *Adv. Energy Mater*. **8**, 1702930 (2018).
115. Boota, M. et al. Activated carbon spheres as a flowable electrode in electrochemical flow capacitors. *J. Electrochem. Soc*. **161**, A1078 (2014).
116. Wang, J., Polleux, J., Lim, J. & Dunn, B. Pseudocapacitive contributions to electrochemical energy storage in $TiO_2$ (anatase) nanoparticles. *J. Phys. Chem. C* **111**, 14925–14931 (2007).
117. Pletcher, D., Peat, R., Robinson, J., Greef, R. & Peter, L. M. Potential sweep techniques and cyclic voltammetry. In *Instrumental Methods in Electrochemistry* (eds Elsevier Science) 178–228 (Woodhead Publishing, 2001).
118. Buczek, S. et al. Rational design of titanium carbide MXene electrode architectures for hybrid capacitive deionization. *Energy Environ. Mater.* **3**, 398–404 (2020).
119. Luo, J. et al. Pillared structure design of MXene with ultralarge interlayer spacing for high-performance lithium-ion capacitors. *ACS Nano* **11**, 2459–2469 (2017).

120. Li, J., Wang, H. & Xiao, X. Intercalation in two-dimensional transition metal carbides and nitrides (MXenes) toward electrochemical capacitor and beyond. *Energy Environ. Mater.* **3**, 306–322 (2020).

121. Alhabeb, M. et al. Guidelines for synthesis and processing of two-dimensional titanium carbide ($Ti_3C_2T_x$ MXene). *Chem. Mater.* **29**, 7633–7644 (2017).

122. Maleski, K., Ren, C. E., Zhao, M. Q., Anasori, B. & Gogotsi, Y. Size-dependent physical and electrochemical properties of two-dimensional MXene flakes. *ACS Appl. Mater. Interfaces* **10**, 24491–24498 (2018).

123. Shuck, C. E. et al. Effect of $Ti_3AlC_2$ MAX phase on structure and properties of resultant $Ti_3C_2T_x$ MXene. *ACS Appl. Nano Mater.* **2**, 3368–3376 (2019).

124. Malm, A. V. & Corbett, J. C. W. Improved dynamic light scattering using an adaptive and statistically driven time resolved treatment of correlation data. *Sci. Rep.* **9**, 13519 (2019).

## ACKNOWLEDGEMENTS

This work is funded in part by the Fulbright Fellowship Program and the Micron School of Materials Science at Boise State University. MXene synthesis and characterization was supported in part by NASA award No. ID-80NSSC17M0029. T.L., L.A., and N.E.M. acknowledge support through the Idaho National Laboratory (INL) Laboratory Directed Research & Development (LDRD) Program under DOE Idaho Operations Office Contract DE-AC07-05ID14517. D.E. acknowledges infrastructure support under DE-NE0008677, joint appointment support under DOE Idaho Operations Office Contract DE-AC07-05ID14517, and career development support from Institutional Development Awards (IDeA) from the National Institute of General Medical Sciences of the National Institutes of Health under Grants #P20GM103408 and P20GM109095. The team acknowledges A. Jerez for his scientific illustrations.

## AUTHOR CONTRIBUTIONS

T.L., N.E.M., Y.G., and D.E. conceived the experimental design. N.E.M. conducted the experiments with assistance from L.A. and C.S. All authors made substantial contributions to data acquisition, analysis, and interpretation. All authors discussed results and co-wrote the final manuscript.

## COMPETING INTERESTS

The authors declare no competing interests.

## ADDITIONAL INFORMATION

**Supplementary information** The online version contains supplementary material available at https://doi.org/10.1038/s41545-022-00164-3.

**Correspondence** and requests for materials should be addressed to Tedd E. Lister or David Estrada.